\documentclass[amsmath,amssymb,floatfix,nopreprintnumbers,nofootinbib,prd,onecolumn,preprint,superscriptaddress]{revtex4}
\usepackage{hyperref}
\usepackage{graphicx}
\usepackage[utf8]{inputenc}
\usepackage[usenames,dvipsnames,svgnames]{xcolor}
\usepackage{amssymb,amsmath,amsthm}
\usepackage{verbatim}

\hypersetup{
  colorlinks,
  linktoc=page,
  linkcolor=blue,
  citecolor=ForestGreen
}

\newcommand{\beq}{\begin{equation}}
\newcommand{\eeq}{\end{equation}}

\newcommand{\gs}{\gamma^\star}
\newcommand{\MSbar}[0]{\overline{\text{MS}}}

\begin{document}

\title{Mass-induced confinement near the sill of the conformal window}

\author{Roman Marcarelli} \email{roman.marcarelli@colorado.edu}

\author{Nicholas Miesch}
\email{nicholas.miesch@colorado.edu}

\author{Ethan T. Neil}
\email{ethan.neil@colorado.edu}

\affiliation{Department of Physics, University of Colorado, Boulder, Colorado 80309, USA}

\begin{abstract}

We revisit standard arguments for hyperscaling of the spectrum when a non-zero fermion mass is introduced to a gauge-fermion theory which is conformal in the infrared limit.  With some general assumptions, we argue that the induced confinement scale will be significantly enhanced near the edge of the conformal to confining transition.  This enhancement can allow for the fermion mass to be arbitrarily small compared to the confinement scale.  This scale separation may allow for apparent spontaneous breaking of chiral symmetry within the conformal window, which may be of interest for construction of dilaton effective field theories in this regime.

\end{abstract}

\maketitle

\pagebreak

\section{Introduction}
	\label{sec:intro}

In the presence of enough fermion degrees of freedom, Yang-Mills gauge theories become scale invariant in the infrared limit.  This effect was noticed following early calculations of the two-loop Yang-Mills $\beta$-function \cite{Caswell:1974gg,Banks:1981nn}.  For a theory with $N_f$ Dirac fermions charged under some representation of the gauge group, the range $N_f^{s} < N_f < N_f^{AF}$ for which a given theory remains infrared-conformal ($N_f > N_f^{s}$) and asymptotically free ($N_f < N_f^{AF}$) is known as the ``conformal window''.  Such theories, and the properties of the conformal field theories (CFTs) that they approach in the far infrared, are of interest in their own right and in the context of composite Higgs models; see \cite{DeGrand:2015zxa,Svetitsky:2017xqk,Witzel:2019jbe,Drach:2020qpj} for recent reviews in the context of lattice calculations investigating the conformal window.

Studying the infrared limit of such a theory requires that no additional mass scales are introduced which would break the emergent conformal symmetry.  In particular, all fermions must be kept massless; for any non-zero fermion mass $m$, at scales $\mu \ll m$ the influence of the fermions on the theory will vanish and the remaining pure-gauge theory will confine.  However, for lattice calculations inside the conformal window, this effect can be a useful feature.  Inducing the confinement of such a theory by working at finite $m_f$ allows the use of traditional lattice methods suited for confining theories, such as spectroscopy of hadronic bound states.  Studying the behavior of the spectrum as a function of $m_f$ (``mass hyperscaling'', to be described further below) can be used to ascertain whether a given theory is indeed within the conformal window, and even determine properties of the infrared-limit CFT (in particular, the anomalous dimension of the mass operator, which dictates the $m$ dependence of the spectrum.)

Standard derivations of mass hyperscaling apply generically to theories inside the conformal window.  Such a generic derivation neglects the effect of how close a given theory is to the ``sill'' of the conformal window at $N_f = N_f^{s}$.  It has been argued previously in the literature \cite{Miransky:1996pd,Miransky:1998dh,Appelquist:1998rb} that as $N_f$ approaches $N_f^{s}$ from \emph{below}, the dynamical fermion mass (an order parameter for spontaneous breaking of chiral symmetry and scale symmetry) should exhibit ``Miransky scaling'', exponential in the inverse distance to the conformal sill:
\beq
m_{\rm dyn} \sim \exp \left( \frac{-C}{\sqrt{N_f^{s} - N_f}} \right).
\eeq
Other authors \cite{Kaplan:2009kr,Gukov:2016tnp} have noted the possibility that the conformal window transition could be an infinite-order transition resembling the Berezinskii-Kosterlitz-Thouless (BKT) phase transition \cite{Berezinsky:1970fr,Kosterlitz:1973xp,Kosterlitz:1974sm} -- we will refer to this scenario as ``BKT-like''.  Miransky scaling of the dynamical mass for $N_f < N_f^{s}$ is also predicted in this scenario.

In this paper, we revisit the standard derivation of mass hyperscaling to consider the effects of varying $N_f$.  In particular, we will argue that as $N_f$ approaches $N_f^{s}$ from \emph{above}, the induced confinement scale $\Lambda_c$ scales as
\beq
\Lambda_c \sim \left( \frac{1}{{N_f - N_f^s}} \right)^{\zeta},
\eeq
where the exponent $\zeta$ is related to the properties of the conformal transition.  For $N_f$ sufficiently close to the sill, this large enhancement of $\Lambda_c$ implies that a parametric range of the theory opens up for which the fermion mass satisfies $m_0 \ll \Lambda_c$.  This raises the intriguing possibility of apparently spontaneous chiral symmetry breaking \emph{within} the conformal window, which could give key insights into extending dilaton effective field theories \cite{Matsuzaki:2013eva,Golterman:2016cdd,Golterman:2016lsd,Hansen:2016fri,Appelquist:2017vyy,Appelquist:2017wcg,Golterman:2018mfm} to $N_f > N_f^{s}$.  Moreover, study of the $N_f$ dependence to extract the $\zeta$ exponent can give direct information on the properties of the conformal transition.

The paper is organized as follows.  In Sec.~\ref{sec:hs_overview}, we review standard arguments for mass hyperscaling, and then provide an alternative description in terms of a mass-dependent renormalization picture, the Gell-Mann-Low renormalization group.  We re-derive standard results in this picture, including a predicted suppression of the confinement scale near $N_f^{AF}$.  Sec.~\ref{sec:hs_sill} contains our main results which concern how this behavior is modified near the conformal sill, where we find a large enhancement of $\Lambda_c$ compared to the fermion mass.  We briefly summarize and discuss the implications of our results in Sec.~\ref{sec:conclusion}.

\section{Mass hyperscaling, revisited} \label{sec:hs_overview}

For the following discussion, we assume that we are working in a Yang-Mills gauge theory with $N_f$ fermion flavors in a single common gauge representation $R$.  The theory is assumed to lie in the conformal window, so that in the limit $\mu \rightarrow 0$ for the massless theory (here $\mu$ is the renormalization-group scale) the gauge coupling $\alpha(\mu) \equiv g(\mu)^2/4\pi$ approaches an infrared fixed point value $\alpha(0) \equiv \alpha^\star$.  The mass anomalous dimension $\gamma(\mu)$ similarly approaches the value $\gamma(0) \equiv \gs$.

\subsection{Confinement and the critical coupling}

An important assumption in the following is that there exists a ``critical coupling'' $\alpha_c$ which determines an approximate confinement scale $\Lambda_c$ through dimensional transmutation, i.e. $\alpha(\Lambda_c) = \alpha_c$.  The existence of such a critical coupling can be justified within certain non-perturbative approximations; for example, solution of Schwinger-Dyson equations in ladder approximation \cite{Cohen:1988sq, Appelquist:1998rb} leads to the prediction of an $\alpha_c$ beyond which solutions exist for a non-zero dynamical fermion mass and chiral symmetry will spontaneously break.  We will not adopt any particular method for estimation of $\alpha_c$; we only need assume that such a critical coupling exists.

By definition, any theory in the conformal window is chirally symmetric and does not confine, and therefore must satisfy $\alpha^\star < \alpha_c$.  Assuming that the conformal transition is continuous, we may alternatively identify $\alpha_c$ as the fixed-point coupling value where exit from the conformal window occurs,
\beq\label{eq:alpha_c_limit}
\alpha_c = \lim_{N_f \searrow N_f^s} \alpha^\star(N_f).
\eeq
In the picture of \cite{Kaplan:2009kr,Gukov:2016tnp} where the transition is BKT-like, the sill of the conformal window corresponds to at least one irrelevant operator $\mathcal{O}$ becoming marginal.  The physical idea is then that even for $N_f > N_f^s$, as the coupling increases to $\alpha_c$, the operator's anomalous dimension $\gamma_\mathcal{O}(\alpha)$ will increase to make it relevant and the theory will confine.  Alternatively, in the BKT-like picture the conformal transition can be interpreted as resulting from the merger of two fixed points, the infrared (IR) fixed point at $\alpha^\star$ and an ultraviolet (UV) fixed point at stronger coupling; the merger occurs at $\alpha^\star = \alpha_c$.

We emphasize that with the assumption of existence of the critical coupling $\alpha_c$, the identification in Eq.~\ref{eq:alpha_c_limit} applies even beyond the BKT-like scenario.  For example, although perturbation theory is not a reliable guide near the bottom of the conformal window,  we may consider the conformal transition predicted by two-loop perturbation theory as a qualitative example. In this case, the predicted value of $\alpha^\star(N_f)$ diverges as $N_f$ decreases (see Appendix~\ref{sec:app_BZ}.)  However, if a given theory confines when $\alpha(\mu) > \alpha_c$ at any scale $\mu$, then the conformal transition in the two-loop scenario must occur at $N_f^s$ such that $\alpha^\star(N_f^s) = \alpha_c$, rather than at the value of $N_f$ for which $\alpha^\star(N_f) \rightarrow \infty$.  

It is possible that the identification Eq.~\ref{eq:alpha_c_limit} may fail under certain scenarios.  For example, in the ``jumping'' scenario of \cite{Sannino:2012wy}, the $\alpha_c$ defined by this limit may be different from the $\alpha_c$ defined through dimensional transmutation of the confinement scale.  In other words, in the jumping scenario we may have $\lim_{N_f \searrow N_f^s} \alpha^\star(N_f) = \alpha_J < \alpha_c$, where the latter is the critical coupling for confinement.  This will not significantly change our results below, which may generally be modified to account for this scenario simply by substituting $\alpha_J$ for $\alpha_c$ in the derivation.  So long as there is not a very large scale separation between $\Lambda_c$ and the scale $\Lambda_J$ defined by $\alpha(\Lambda_J) = \alpha_J$, the jumping scenario should exhibit the same enhancement of $\Lambda_c$ as the bottom of the conformal window is approached.

\begin{figure}
  \centering
  \includegraphics[width=\textwidth]{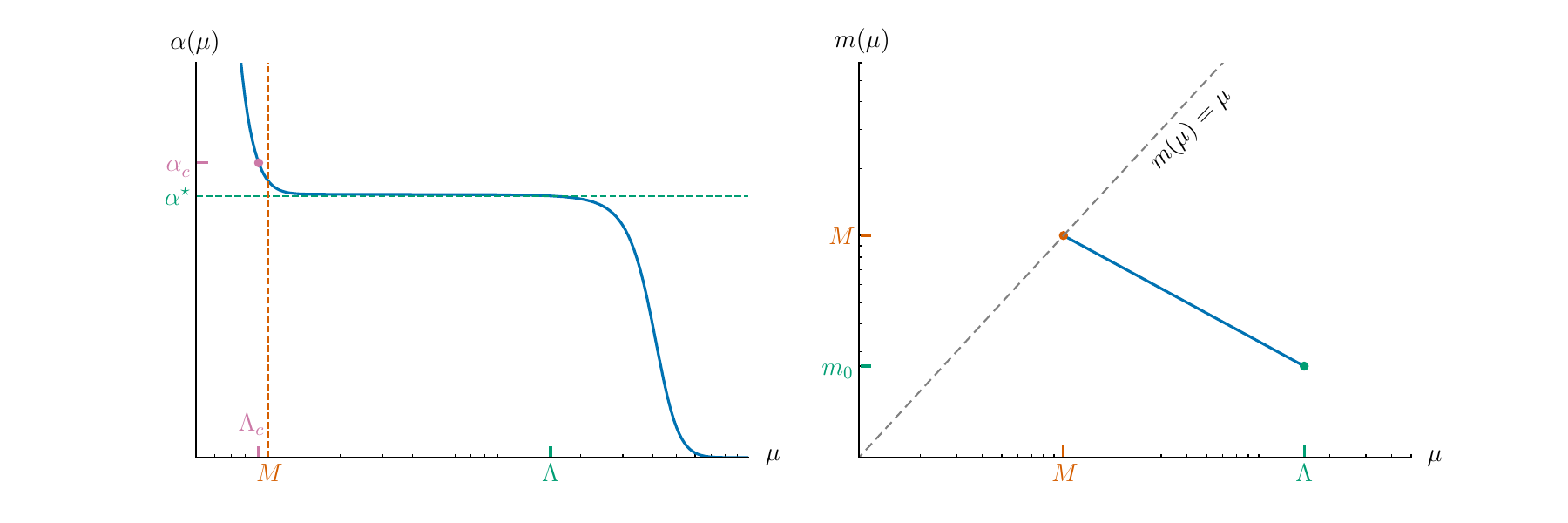} \\
    \caption{\label{fig:hyperscale} Sketch of the scale dependence of the running coupling $\alpha(\mu)$ and the running fermion mass $m(\mu)$ in the standard mass hyperscaling scenario, as described in the text.}
\end{figure}

\subsection{Mass hyperscaling: the standard picture} \label{ssec:standard_hs}

We begin by outlining the standard mass hyperscaling scenario.  A sketch of the derivation is given in Fig.~\ref{fig:hyperscale}.  First, we identify an ultraviolet scale $\Lambda$ at which a ``seed mass'' $m_0 = m(\Lambda)$ is introduced into the theory.  The scale $\Lambda$ is chosen so that it is within the regime where the theory is approximately scale-invariant, i.e.~$\alpha(\Lambda) \approx \alpha^\star$ and thus $\gamma(\Lambda) \approx \gs$.  (Because the beta function $d\alpha / d \mu$ is vanishing in the infrared limit, it is always possible to choose such a $\Lambda$.)  At lower scales, the mass evolves according to its anomalous dimension as
\beq
m(\mu) = m_0 \left( \frac{\Lambda}{\mu} \right)^{\gs}.
\eeq
As the RG scale $\mu$ decreases, the mass increases, until eventually they are comparable; we define the energy scale at which they are equal to be $M$, i.e.
\beq \label{eq:hs_M}
M \equiv m(M).
\eeq
Solving this equation for $M$ using the RG evolution of the mass gives the relationship between $M$ and the seed mass $m_0$ to be
\beq
M = \Lambda^{\frac{\gs}{1+\gs}} m_0{}^{\frac{1}{1+\gs}}.
\eeq
At scales $\mu < M$, the fermions no longer have negligible mass, and in particular as $\mu \ll M$ they can be integrated out of the theory \cite{Appelquist:1974tg}, leaving a pure-gauge Yang-Mills theory.  The $\beta$-function of such a theory will be large and negative, quickly driving $\alpha \rightarrow \alpha_c$ and triggering confinement and chiral symmetry breaking.  Assuming that this happens rapidly as soon as the fermion mass is no longer negligible, we have the result for the confinement scale
\beq \label{eq:standard-hs}
\Lambda_c \approx M \propto m_0{}^{\frac{1}{1+\gs}}.
\eeq
This is the mass hyperscaling relation \cite{DelDebbio:2010ze,DelDebbio:2010jy,Appelquist:2011dp,DeGrand:2011cu,Cheng:2013xha}; it predicts that all hadronic quantities will scale as a power law of the fermion mass.  This power-law scaling has been observed in a number of lattice simulations \cite{Bursa:2011ru,Appelquist:2011dp,DeGrand:2011cu,Cheng:2013xha,DelDebbio:2015byq,LatKMI:2015ppz,Bergner:2016hip,Hasenfratz:2017hdd}.

\subsection{Running coupling in the Gell-Mann-Low renormalization group \label{ssec:gml}}

Although the derivation above does not make explicit reference to a renormalization scheme, it is implicitly carried out in a Callan--Symanzik (CS) renormalization group scheme (such as $\MSbar$) where both the coupling $\alpha(\mu)$ and mass parameter $m(\mu)$ are taken to depend directly on the RG scale $\mu$.  For our analysis to follow, we will instead find it convenient to adopt a renormalization scheme in which the fermion mass $m$ is fixed equal to the physical or pole mass $m_P$.  Specifically, we will adopt the Gell-Mann--Low (GML) renormalization group picture. \cite{Gell-Mann:1954yli} Although most modern renormalization procedures are performed in the CS renormalization group, we find that the GML renormalization group is advantageous for our purposes due to the explicit dependence of $\alpha$ on the particle mass.  This is in contrast to $\MSbar$, in which $\alpha$ is taken to be independent of fermion mass, with $m$ only influencing the running through decoupling of the fermions at mass threshold.  Our results should of course be independent of renormalization scheme, but the derivation is clearer in GML.  For more information, the relationship between these renormalization groups is described in Refs. \cite{Wilson:1974mb,Acharya:1978xz, Higashijima:1980fm, Brodsky:1998mf,Brodsky:1999fr}.

An advantage of using the GML renormalization group is that rather than having mass run with renormalization scale, the mass is fixed to be the pole of the propagator, $m_P$, with the trade-off that the running of $\alpha$ now depends on both the renormalization scale $Q$ and the mass $m_P$. Hence, the analogue of the $\beta$-function, the Gell-Mann--Low $\Psi$ function, is not only a function of $\alpha$ but also of $Q^2/m_P^2$:
\begin{align}
    \frac{d\alpha}{d\log Q^2} &= \Psi(\alpha, Q^2/m_P^2) = -\frac{\alpha^2}{4\pi}\sum_{k=0}^{\infty}\psi^{(k)}(Q^2/m_P^2)\left(\frac{\alpha}{4\pi}\right)^k
\end{align}
In order to investigate the confinement scale, it is useful to work in a scheme in which the coupling constant is defined in terms of the QCD static quark potential \cite{Lepage:1992xa} (we will refer to this as the $V$-scheme), defining the coupling such that
\begin{align}
    V(Q^2) = -C_F\frac{4\pi \alpha(Q^2)}{Q^2}
\end{align}
where $C_F$ is the Casimir invariant in the fundamental representation.

There is a useful and direct connection between the pole mass $m_P$ and the mass scale $M$ introduced in Eq.~\ref{eq:hs_M} above: they are directly proportional.  The pole mass itself is physical and scheme-independent; on the other hand, $M$ is defined through $M = m(M)$ based on a running-mass renormalization scheme.  As discussed in \cite{Hoang:2017btd}, the relationship between the pole mass and the mass scale $M$ in the $\MSbar$ scheme can be written as an asymptotic expansion in the coupling $\alpha$, including threshold effects.  The details are unimportant for our argument; what matters is that $M \propto m_P$, with no additional scale dependence.

To finish the connection to hyperscaling using the Gell-Mann--Low renormalization group, we observe that due to the dependence of $\alpha$ on $Q^2/m_P^2$, the condition $\alpha(Q=\Lambda_c) = \alpha_c$ immediately implies the relation
\begin{equation}
\Lambda_c^2 = m_P^2 \alpha^{-1}(\alpha_c).
\end{equation}
The relationship $\Lambda_c \propto M$ is thus recovered naturally on dimensional grounds; in a mass-deformed conformal theory, we may further identify $M \propto m_0^{1/(1+\gs)}$ to recover the standard hyperscaling relation.  Moreover, working in this scheme allows us to directly examine the factor of proportionality between $\Lambda_c$ and $m_P$, the function $\alpha^{-1}(\alpha_c)$, and in particular how it depends on the number of fermions present.

\subsection{Mass hyperscaling near the top of the conformal window}

The standard hyperscaling derivation of Sec.~\ref{ssec:standard_hs} applies universally throughout the conformal window, making no reference to the value of $\alpha^\star$.  This is because the running needed to induce confinement from $\alpha^\star$ to $\alpha_c$ is ignored, presuming that it is a small effect.  We can attempt to improve the derivation and elucidate the dependence on $N_f$ by taking this running into account.

We begin by studying the Caswell-Banks-Zaks limit \cite{Caswell:1974gg, Banks:1981nn}, with $N_f$ close to $N_f^{AF}$, so that $\alpha^\star$ can be reliably obtained with two-loop perturbation theory.  The qualitative behavior that we expect in this limit was pointed out by Miransky \cite{Miransky:1998dh}: as $\alpha^\star$ becomes very small, the effect of running from $\alpha^\star$ to $\alpha_c$ opens up an exponentially large scale separation between $M$ and $\Lambda_c$ (so that, for example, the masses of glueballs are predicted to be highly suppressed.)

This scale separation effect can be obtained in a running-mass scheme, as was done in \cite{Miransky:1998dh}.  We will study this effect using the Gell-Mann--Low renormalization group, which will allow us to obtain a formula for the scale separation and will also serve to validate our methods.  We will work in the large-$N_c$ limit, which will simplify the algebra somewhat, although we expect the qualitative result to hold at finite $N_c$.  As a further check, this result can also be readily obtained in $\MSbar$; see Appendix~\ref{sec:app_BZ}.

In the $V$-scheme, the running of the coupling in the large $N_c$ limit of an $SU(N_c)$ gauge theory with $N_f = n_f N_c$ fermions of mass $m_P$ in two-loop perturbation theory is given by the equation
\begin{equation}
    \frac{d\lambda}{d\log Q^2} = -\psi^{(0)}(Q^2/m_P^2)\lambda^2 - \psi^{(1)}(Q^2/m_P^2)\lambda^3 + \cdots \label{eq:two_loop_psi}
\end{equation}
where we are using $\lambda = \frac{N_c}{4\pi}\alpha$ and the first two coefficients are given by \cite{Brodsky:1999fr}
\begin{align}
    \psi^{(0)}(x) = \frac{11}{3}-\frac{2}{3}n_f + \frac{2}{3}n_f f^{(0)}(x),
\intertext{and}
    \psi^{(1)}(x) = \frac{34}{3}-\frac{13}{3}n_f + \frac{13}{3}n_f f^{(1)}(x),
\end{align}
where $x = Q^2/m_P^2$.  The functions $f^{(i)}(x)$ satisfy $\lim_{x\rightarrow 0}f^{(i)}(x) = 1$ and $\lim_{x\rightarrow\infty}f^{(i)}(x) = 0$, reflecting that the theory effectively has zero fermions at $Q\ll m_P$, and $n_f N_c$ fermions at $Q\gg m_ P$. The function $f^{(0)}(x)$ is determined exactly in Ref.~\cite{Brodsky:1998mf} as 
\begin{equation}
    f^{(0)}(x) = \frac{6}{x}\left[1 - \frac{4}{x}\sqrt{\frac{x}{x+4}}\tanh^{-1}\left(\sqrt{\frac{x}{x+4}}\right)\right]
\end{equation}
while $f^{(1)}(x)$ is determined numerically in Ref. \cite{Brodsky:1999fr}.

We will consider a perturbative expansion about $\delta n_f^{AF} \equiv n_f^{AF} - n_f = 11/2 - n_f$, where the infrared fixed-point coupling (in the massless limit $x \rightarrow \infty$) is
\begin{equation}
    \lambda^\star = \frac{11-2n_f}{13n_f - 34} \approx \frac{4}{75}\delta n_f^{AF}.\label{eq:alphaIRFP}
\end{equation}
Since $\lambda^\star = {\cal O}(\delta n_f^{AF})$, we expect $\lambda = {\cal O}(\delta n_f^{AF})$ until $x \ll 1$, and hence we can expand to first order in $\lambda$. Doing so yields a separable differential equation
\begin{equation}
    \frac{d\lambda}{d\log x} = - \frac{11}{3}\lambda^2f^{(0)}(x).
\end{equation}
Solving this equation requires us to specify an initial condition; in line with the hyperscaling derivation presented in Sec.~\ref{ssec:standard_hs}, we want to choose an ultraviolet scale $\Lambda$ where $\lambda(\Lambda) \approx \lambda^\star$, so that the (small) running of the coupling between $0$ and $\lambda^\star$ is not relevant.  The simplest possibility is to take the limit $\Lambda \rightarrow \infty$ (equivalent to working in the limit $m_P \ll \Lambda$.)  To do so, we define
\begin{align}
    F(x) &= \int_{x}^{\infty}\frac{f^{(0)}(x')}{x'}dx'\nonumber\\
    &= \frac{4}{x}-\log{x} + \frac{2(x-2)}{x}\sqrt{\frac{x+4}{x}}\tanh^{-1}\left(\sqrt{\frac{x}{x+4}}\right)
\end{align}
and then we can arrive at a solution to $\lambda$ which corresponds to $\lambda(\infty) = \lambda^\star$. We then have
\begin{align}
    \lambda &= \frac{12 \delta n_f^{AF}}{225 - 44\delta n_f^{AF} F(x)}\label{eq:GML_alpha}
\end{align}
It can be verified numerically that for small $\delta n_f^{AF}$, the inversion of Eq.~\ref{eq:GML_alpha} corresponds to a small value of $x$. Hence, we can expand Eq.~\ref{eq:GML_alpha} for small $x$ and solve for the scale at which $\delta \lambda = \delta \lambda_c$. Doing so, we find a confinement scale
\begin{equation}
      \Lambda_c^2 = m_P^2 \exp\left(\frac{5}{3}+\frac{3}{11\lambda_c} - \frac{225}{44\delta n_f^{AF}}\right)
\end{equation}
indicating that along with the expected scaling proportional to the pole mass $m_P$, the confinement scales with $\delta n_f^{AF}$ according to
\begin{align} \label{eq: BanksZaksGML}
    \frac{\Lambda_c}{m_P} \sim \exp\left(-\frac{225}{88\delta n_f^{AF}}\right),
\end{align}
so that the confinement scale will be exponentially suppressed. This reproduces the predicted scaling of Ref.~\cite{Miransky:1998dh}, including the numerical prefactor which comes from perturbation theory (see Appendix~\ref{sec:app_BZ}.)  It is further noted in Ref.~\cite{Miransky:1998dh} that an observable consequence of this suppression could be that glueball states (whose mass will be set by $\Lambda_c$) become exponentially light compared to fermionic bound states (whose mass will be dominated by $m_P$ if $m_P \gg \Lambda_c$.)


\section{Mass hyperscaling approaching the conformal sill} \label{sec:hs_sill}

Next, we turn to the case where $N_f \searrow N_f^s$.  In this case, we have $\lambda^\star \approx \lambda_c$, so that we expect that the running between these two scales will be small indeed (as assumed in the standard hyperscaling derivation.)  However, as $\lambda_c - \lambda^\star \rightarrow 0$, another formerly negligible effect can become important: the influence of the fermion mass itself on  the running of $\lambda(\mu)$.  We know that this influence is important, because the hyperscaling derivation relies on decoupling; that is, the $\beta$-function and thus $\lambda(\mu)$ changes dramatically for $\mu \gg M$ and for $\mu \ll M$, becoming the $\beta$-function for an $N_f = 0$ pure-gauge theory in the latter case.

\begin{figure}
  \centering
  \includegraphics[width=\textwidth]{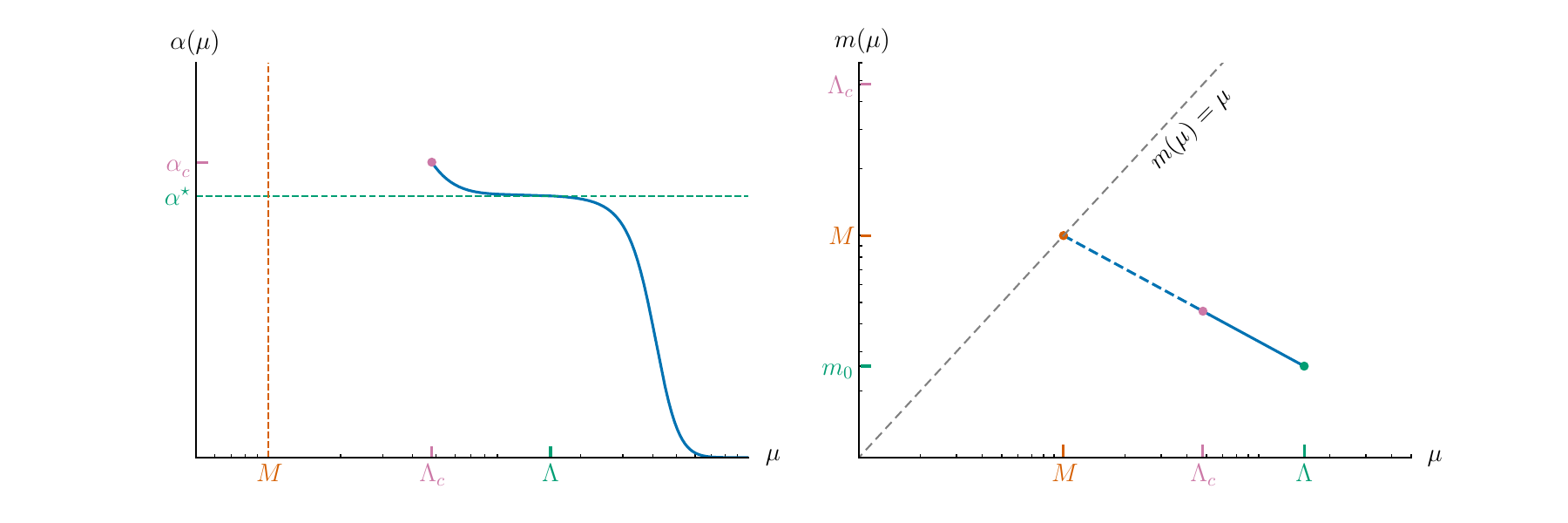} \\
    \caption{\label{fig:hyperscale-sill} Sketch of how the standard hyperscaling argument breaks down near the sill of the conformal window. }
\end{figure}

When $\lambda^\star$ is close enough to $\lambda_c$, a relatively small change in the coupling due to the introduction of a fermion mass should be sufficient to cause the theory to confine.  Below, we will argue that with careful consideration of the order of limits, there exists a regime where the confinement scale is arbitrarily large compared to the fermion mass, $\Lambda_c \gg m$.  We begin with an existence argument on general grounds, and then repeat the derivation with a specific form for $\Psi$-function.
The qualitative situation is depicted in Fig.~\ref{fig:hyperscale-sill}.

\subsection{Existence argument}\label{ssec:existence}

Consider a theory inside the conformal window with $\lambda^* = \lambda_c - \epsilon_c$ and $\epsilon_c \ll 1$.  In the large-$N_c$ limit, we can adjust $n_f - n_f^s$ continuously to satisfy this condition for any $\epsilon_c$.  To study the effect of a non-zero fermion mass, we will denote the difference between the running coupling in the massless theory and in the massive theory by

\beq
\Delta \lambda(Q^2) \equiv \lambda_{m = m_P}(Q^2) - \lambda_{m = 0}(Q^2).
\eeq

This comparison requires us to hold everything else fixed, aside from the fermion mass.  To do so explicitly, we can identify an energy scale $\Lambda$ such that $\lambda(\Lambda^2) = \lambda^\star - \epsilon_\Lambda$ for some infinitesimal $\epsilon_\Lambda$, but also require $m_P \ll \Lambda$ so that the effect of the mass on the running near $\Lambda$ is negligible.  Since the $\beta$-function in the massless theory has a zero, this separation of scales can be arbitrarily large.  (Roughly speaking, $\Lambda$ is on the infrared ``plateau'' region of the running coupling where the $\beta$-function is approximately zero.)  Because the effect of the fermions is always one of screening, adding a fermion mass will always increase the $\beta$-function and therefore $\Delta \lambda(Q^2)$ is strictly positive. 

Suppose we evaluate $\Delta \lambda(Q^2)$ at some value of $Q\gg m_P$.  By construction, $\Delta\lambda(Q)$ should go to 0 as $m_P$ does, so we can write
\beq \label{eq:mDeltaalphaExpansion}
\Delta \lambda(Q^2) = (a + \mathcal{O}(\epsilon_c)) \frac{m_P^2}{Q^2} + \mathcal{O}\left( \frac{m_P^4}{Q^4}\right) + \mathcal{O}\left(\frac{m_P^2}{\Lambda^2} \right)
\eeq
where we neglect terms containing $m_P / \Lambda$ because we have already ensured that this ratio is small.  

Keeping full generality, we can also rewrite $\lambda_{m = 0}(Q^2)$ for $Q<\Lambda$ in the following form to highlight its $\epsilon_\Lambda$ dependence:
\beq \label{eq:alphaMasslessExpansion}
\lambda_{m=0}(Q^2) = \lambda^\star - (b(\epsilon_\Lambda, Q) + \mathcal{O}(\epsilon_c)){\epsilon_\Lambda} - ... 
\eeq
Here the function $b$ is not assumed to be analytic, however we do know it must be between 1 and 0 for all $Q<\Lambda$ because of the definition of $\epsilon_\Lambda$ and the fact that $\lambda(Q^2)$ is monotonic.

Now we can use these relations to manufacture a $\Lambda_c$ in the regime we are hoping for.  Suppose a new arbitrary mass scale $Q_0$ is \textit{defined} as the solution to the equation
\beq \label{eq:LambdaCExpansion}
a \frac{m_P^2}{Q_0^2}=\epsilon_c + {b}({\epsilon_\Lambda},Q_0)\epsilon_\Lambda
\eeq
It is not difficult to see that since $\epsilon_c$ and $\epsilon_\Lambda$ are assumed to be very small, we have $\frac{m_P^2}{Q_0^2}\ll 1$ as well. (For an explicit solution verifying this, see Eq. \ref{eq:recursiveFormula}.)  Equations \ref{eq:mDeltaalphaExpansion} and \ref{eq:alphaMasslessExpansion} can then be used to rewrite the above expression as
\begin{align}
    \Delta\lambda(Q_0^2)&=\epsilon_c + (\lambda^*-\lambda_{m=0}(Q_0^2)) \\
    \implies \lambda_{m = m_P}(Q_0^2) - \lambda_{m = 0}(Q_0^2)&=\lambda_c-\lambda_{m=0}(Q_0^2) \\
    \implies \lambda_{m = m_P}(Q_0^2)&=\lambda_c \\
    \implies Q_0&=\Lambda_c.
\end{align}
Eq. \ref{eq:LambdaCExpansion} can now be solved to find general solutions for $\Lambda_c$.

To make this process more straightforward, we can use the parametric control over $\epsilon_\Lambda$ we have access to. No assumptions were made about $\epsilon_c$ and $\epsilon_\Lambda$ other than that they are small, so we are free to take relative limits of them without losing the existence of $\Lambda_c$.  The region we are most interested in occurs when $\epsilon_c\gg \epsilon_\Lambda$.  In this case, one obtains

\beq \label{eq:nfRegimeLambdaC}
    \epsilon_c = a \frac{m_P^2}{\Lambda_c^2} \implies
    \Lambda_c\propto\frac{m_P}{\sqrt{\epsilon_c}}.
\eeq

This is a very general result, and it establishes the existence of a regime for which $m_P \ll \Lambda_c$.  To make it more practically useful, we can considering the relationship between $\epsilon_c$ and $n_f$.  We know that $\epsilon_c \rightarrow 0$ as $n_f \rightarrow n_f^s$ by definition.  Assuming the relationship between the two quantities is analytic, we may write at small $\epsilon_c$
\beq
\delta n_f = n' \epsilon_c^k + \mathcal{O}(\epsilon_c^{k+1})\label{eq:nfexpansion}
\eeq
where $k$ is an integer corresponding to the first non-zero order in the expansion.  (This form and the integer $k$ can also be derived by considering an expansion of the $\beta$ or $\Psi$ function about $\lambda_c$ - see Sec~\ref{ssec:alpha_star_exp} below.)  From Eq. \ref{eq:nfRegimeLambdaC}, the physical confinement scale can then be expressed scheme-independently as
\beq
\Lambda_c\propto \frac{m_P}{(\delta n_f)^{\zeta}},
\eeq
where the exponent is defined as $\zeta\equiv\frac{1}{2k}$.

The value of the exponent $k$ is closely related to the nature of the conformal window transition.  For example, suppose that the transition is ``BKT-like'', with two fixed-point solutions merging at $n_f = n_f^s$.  Then recalling the definition $\epsilon_c = \lambda_c - \lambda^\star$, for any $n_f > n_f^s$ we expect to find two solutions for $\epsilon_c$, while at $n_f < n_f^s$ there should be no solutions.  This behavior is consistent with only even terms appearing in the expansion Eq.~\ref{eq:nfexpansion}, with the simplest possibility being $k = 2$.  

We may contrast this with a scenario in which the transition at the sill happens abruptly. An example of this is a situation in which the two-loop $\Psi$-function (Eq. \ref{eq:two_loop_psi}) is exact, so that the sill is crossed at the value of $n_f^s$ for which $\lambda^\star(n_f^s)  = \lambda_c$, with $\lambda^*$ given by Eq.~\ref{eq:alphaIRFP}. Expanding about $n_f = n_f^s$, we have $\epsilon_c = \lambda_c-\lambda^\star = 75\delta n_f/(13n_f^s - 34)^2$, implying $k=1$. We will hence refer to $k=1$ as a ``PT-like'' scenario, although we recognize that there are other scenarios in which one may have $\epsilon_c \propto \delta n_f$. Measurement of the exponent $\zeta$ allows us to determine the value of $k$ and shed light on the nature of the conformal window transition.

Although the results above with $\epsilon_c \gg \epsilon_\Lambda$ are of primary interest to us, it is worth noting that the opposite limit $\epsilon_\Lambda \gg \epsilon_c$ leads to a secondary region of parameter space for Eq.~\ref{eq:LambdaCExpansion} with different properties.  In this case, we have
\beq \label{eq:epsilonLambdaRegime}
    a \frac{m_P^2}{\Lambda_c^2} = {b}({\epsilon_\Lambda},\Lambda_c)\epsilon_\Lambda
\eeq

Recall that $b$ is totally $\delta n_f$-independent through its definition as the 0th order term in an $\epsilon_c$-expansion.  Therefore, this solution for $\Lambda_c$ has no $\delta n_f$ dependence.  Thus, if we hold $\epsilon_\Lambda$ fixed and attempt to tune $n_f \rightarrow 0$, we will eventually enter this solution regime and the dependence of $\Lambda_c$ on $\delta n_f$ will vanish again.  (This may always be avoided, however, by taking $\epsilon_\Lambda$ to always be small enough that the $\epsilon_c$ term is never negligible.)

The $\Psi$-function based arguments in the next section do provide insight into the explicit form of $b(\epsilon_\Lambda,\Lambda_c)$, allowing us to actually solve Eq. \ref{eq:epsilonLambdaRegime}.  We obtain in Eq. \ref{eq:dalpha} that for some positive real number $p$,
\beq
    {b}({\epsilon_\Lambda},Q) = b_0\frac{Q^{2p}}{\Lambda^{2p}}+O\left(\frac{Q^{4p}}{\Lambda^{4p}}\right)
\eeq
\beq
    \implies a \frac{m_P^2}{\Lambda_c^2} =b_0 \epsilon_\Lambda\frac{\Lambda_c^{2p}}{\Lambda^{2p}}
\eeq
\beq
    \implies\Lambda_c\propto \Lambda \left(\frac{1}{\sqrt{\epsilon_\Lambda}}\frac{m_P}{\Lambda}\right)^\frac{1}{p+1}.
\eeq

If $p$ is small, as it is in perturbation theory, then $\Lambda_c$ is once again just governed by $m$:
\beq
    \Lambda_c\propto \frac{m_P}{\sqrt{\epsilon_\Lambda}}.\label{eq:epslambda}
\eeq

There is an interesting similarity between Eq.~\ref{eq:epslambda} and Eq.~\ref{eq:nfRegimeLambdaC}; in both cases the enhancement of the confinement scale is present, with $\Lambda_c$ determined by the inverse square root of an infinitesimal difference in $\lambda$.  The key change from the first equation to the second is the complete lack of $\delta n_f$ dependence.

\subsection{Expansion of the $\Psi$ function \label{ssec:alpha_star_exp}}

Having established the existence of a regime where $m_P \ll \Lambda_c$ with characteristic scaling of the confinement scale with $\delta n_f = n_f - n_f^s$ by the general argument above, we now turn to an alternative analysis based on study of the $\Psi$-function in the same regime.  In this section we assume $m_P^2 / \Lambda^2 \ll 1$, restricting us to the light-mass scaling regime found above.

To study the $\delta n_f$ dependence of $\Lambda_c$ in the $\delta n_f \rightarrow 0$ regime, we consider a $\Psi$-function of the form
\begin{equation}
\frac{d\lambda}{d\log Q^2} = -f(\lambda, m_P^2/Q^2) \lambda^2 \left[ (\lambda_c - \lambda)^{k} - A\delta n_f + g(\lambda, m_P^2/Q^2) \right]\label{eq:general_running}
\end{equation}
where $k$ is a positive integer, $g(\lambda, 0) = 0$, and $f(\lambda, 0)$ has no zeros in the region of interest.  This form is expected to capture the most general possible structure of the $\Psi$-function in this regime, near $\lambda^\star \approx \lambda_c$.  When $m_P = 0$, there is a fixed point at $\lambda^\star = \lambda_c - (A\delta n_f)^{1/k}$, so $\epsilon_c = (A\delta n_f)^{1/k}$.  For even $k$, it can be seen that $\lambda^\star$ becomes imaginary for $n_f < n_f^s$. This is similar to the BKT-like scenario suggested by Ref.~\cite{Kaplan:2009kr}, with the simplest possible realization of this effect being at $k = 2$. For odd $k$, $\lambda^\star > \lambda_c$ for $n_f < n_f^s$, so the theory confines before the IR fixed point is reached.  We briefly note that one can consider a slightly more exotic form where the term containing $\delta n_f$ is given by $A(\delta n_f)^{2m+1}$ for any positive integer $m$. Though this recovers the same phenomenological aspects of Eq.~\ref{eq:general_running}  for odd and even $k$, such a structure for the leading term in $\delta n_f$ would require miraculous cancellations at all orders in perturbation theory, so we omit it for simplicity.

To analyze the scaling behavior of $\Lambda_c$ with $\delta n_f$, we will obtain a solution for $\lambda(Q^2)$ in the regime of interest.  Invoking our general argument (see Fig.~\ref{fig:hyperscale-sill}), we anticipate that for $\delta n_f \ll 1$ the theory will confine long before the scale of the fermion mass $m_P$ is reached. As a result, we may expand in $m_P^2/Q^2$. Alternatively, we may begin by assuming $m_P^2 \ll Q^2$, which will limit the range of validity of our solution; if we find that our solution remains valid at $Q^2 = \Lambda_c^2$, then we have a self-consistent result.

Since we are only concerned with the running very close to the fixed-point coupling (since $\lambda_c$ is close to $\lambda^\star$), we will expand $\lambda = \lambda^\star + \delta\lambda$ and work in terms of $\delta \lambda$. Doing so yields
\begin{align}
    \frac{d(\delta \lambda)}{d\log Q^2} &= -f(\lambda^\star, 0)\lambda^{\star 2}\left[(\epsilon_c-\delta\lambda)^k - \epsilon_c^k + g^{(0, 1)}(\lambda^\star, 0)\frac{m_P^2}{Q^2}\right].\label{eq:diffeqdalpha}
\end{align}
For $k > 1$, perturbative analysis reveals that the first two terms in the differential equation are always small compared to the second, since $\epsilon_c \sim \delta\lambda \propto m_P^2/Q^2$, and $(m_P^2/Q^2)^k \ll m_P^2/Q^2$. For $k = 1$, this is no longer the case, but the differential equation can be solved exactly. For brevity, we keep the linear term in $\delta \lambda$ when solving for $k > 1$, even though it is smaller than $m_P^2/Q^2$ by a factor of $\epsilon_c^{k-1}$. Doing so, we can solve for any $k \geq 1$, finding that with initial condition $\delta\lambda(\Lambda^2) = -\epsilon_\Lambda$, Eq. \ref{eq:diffeqdalpha} has solution
\begin{align}
    \delta\lambda &\approx \frac{f(\lambda^\star,0)\lambda^{\star2}g^{(0,1)}(\lambda^\star,0)}{1+k\epsilon_c^{k-1}f(\lambda^\star,0)\lambda^{\star2}}\left[\frac{m_P^2}{Q^2}-\frac{m_P^2}{\Lambda^2}\left(\frac{Q^2}{\Lambda^2}\right)^{k\epsilon_c^{k-1}f(\lambda^\star,0)\lambda^{\star2}}\right] - \epsilon_\Lambda \left(\frac{Q^2}{\Lambda^2}\right)^{k\epsilon_c^{k-1}f(\lambda^\star,0)\lambda^{\star2}}\\
    &\approx \frac{f(\lambda^\star,0)\lambda^{\star2}g^{(0,1)}(\lambda^\star,0)}{1+k\epsilon_c^{k-1}f(\lambda^\star,0)\lambda^{\star2}}\frac{m_P^2}{Q^2} - \epsilon_\Lambda \left(\frac{Q^2}{\Lambda^2}\right)^{k\epsilon_c^{k-1}f(\lambda^\star,0)\lambda^{\star2}}\label{eq:dalpha}
\end{align}
We note that the dependence on $Q^2/\Lambda^2$ matches that found for $g-g^*$ in Ref.~\cite{DelDebbio:2013qta}, and the additional mass dependence is due to us working in the Gell-Mann-Low renormalization group. Our scaling results appear qualitatively different than those of Ref.~\cite{DelDebbio:2013qta}, but this is because we have taken the limit $n_f \rightarrow n_f^s$ in addition to $\alpha \rightarrow \alpha^*$. We can now use Eq.~\ref{eq:dalpha} to solve for the confinement scale, which occurs at $\lambda = \lambda_c$ ($\delta \lambda =\epsilon_c =  (A\delta n_f)^{1/k}$). Substituting $\delta\lambda = \epsilon_c$ and $Q = \Lambda_c$ into Eq.~\ref{eq:dalpha} and rearranging yields
\begin{align}
    \Lambda_c^2 \left[1 + \frac{\epsilon_\Lambda}{\epsilon_c}\left(\frac{\Lambda_c^2}{\Lambda^2}\right)^{k\epsilon_c^{k-1}f(\lambda^\star,0)\lambda^{\star2}}\right] = \frac{f(\lambda^\star,0)\lambda^{\star2}g^{(0,1)}(\lambda^\star,0)}{1+k\epsilon_c^{k-1}f(\lambda^\star,0)\lambda^{\star2}}\frac{m_P^2}{\epsilon_c} \label{eq:recursiveFormula}
\end{align}
\begin{figure}[t!]
    \centering
    \includegraphics[width=0.8\linewidth]{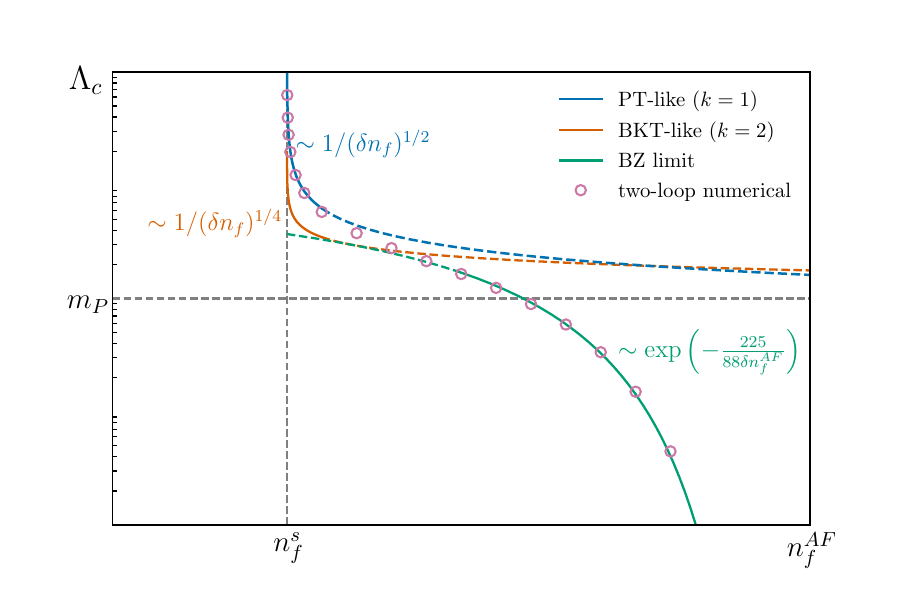}
    \caption{Illustrative sketch showing, within the conformal window, the scale of confinement $\Lambda_c$ vs. the number of fermions $n_f = N_f/N_c$ in the large $N_c$ limit for two confinement hypotheses. The plot demonstrates the qualitative scaling of $\Lambda_c$ vs. $n_f$ in the $k=1$ (PT-like) scenario, and the $k=2$ (BKT-like) scenario. To produce the plots, we have arbitrarily taken $n_f^s = 4$ (corresponding to $\lambda_c = 1/6$ in the two-loop case).  Loss of asymptotic freedom occurs at $n_f^{AF} = 11/2$.  The numerical results are obtained by solving Eq.~\ref{eq:two_loop_psi} for $\lambda$ and solving $\lambda(Q^2) = \lambda_c$ for $Q$. The initial condition was taken to be $\lambda(10^3m_P) = \lambda^\star(n_f)$ for $n_f > 5.2$ and $\lambda(10^6m_P) = \lambda^\star(n_f)$ for $n_f \leq 5.2$.} \label{fig:confinement_scale}
\end{figure}
If $\epsilon_\Lambda \ll \epsilon_c$, the running of $\lambda$ is primarily due to the region $\lambda^\star < \lambda < \lambda_c$, and $\Lambda_c$ scales according to
\begin{align}
    \Lambda_c \sim \frac{m_P}{\sqrt{\epsilon_c}} \sim \frac{m_P}{(\delta n_f)^{1/2k}}.
\end{align}
Hence, in addition to hyperscaling, there is an enhancement proportional to some inverse power of $\delta n_f$.  $k=1$ corresponds to a scenario like two-loop perturbation theory, where the confinement happens abruptly at some $n_f^s$ before $\lambda^\star(n_f^s)$ diverges and the form of the $\Psi$ function is always linear (PT-like).  $k=2$ corresponds to a BKT-like scenario, as described in Ref.~\cite{Kaplan:2009kr}, where an IR and UV fixed point meet at $\lambda^*(n_f^s) = \lambda_c$, marking the end of the conformal window (BKT-like).  In the latter case, the fixed-point merger results in a second-order zero precisely at the end of the conformal window. The behavior of $\Lambda_c$ over the range $n_f^s < n_f < n_f^{AF}$ is depicted in Fig.~\ref{fig:confinement_scale} for PT-like ($k=1$) and BKT-like ($k=2$) confinement. For $n_f \approx n_f^{AF}$, the two confinement mechanisms are indistinguishable, but for $n_f \rightarrow n_f^s$, differences in scaling are apparent. 

\begin{figure}[t!]
    \centering
    \includegraphics[width=0.8\linewidth]{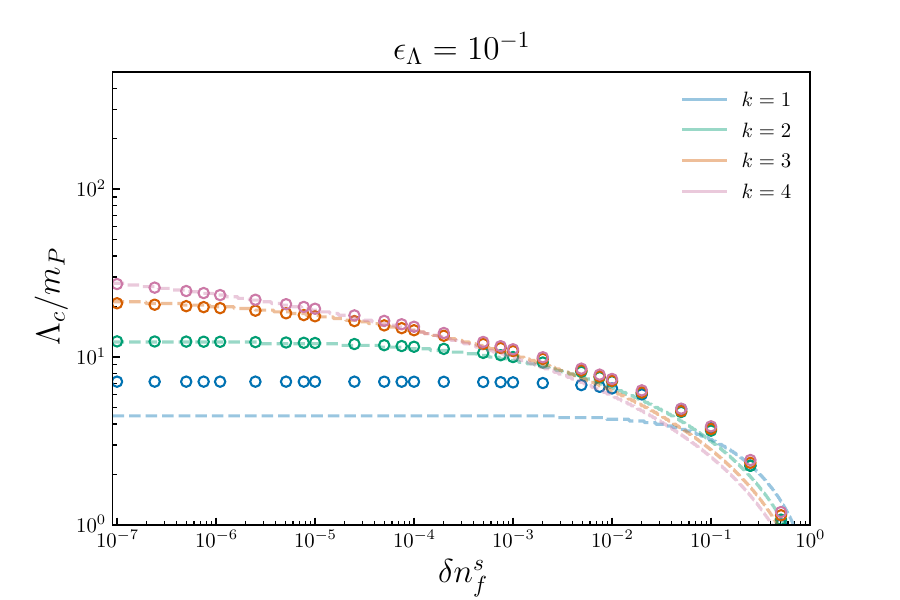}
    \includegraphics[width=0.8\linewidth]{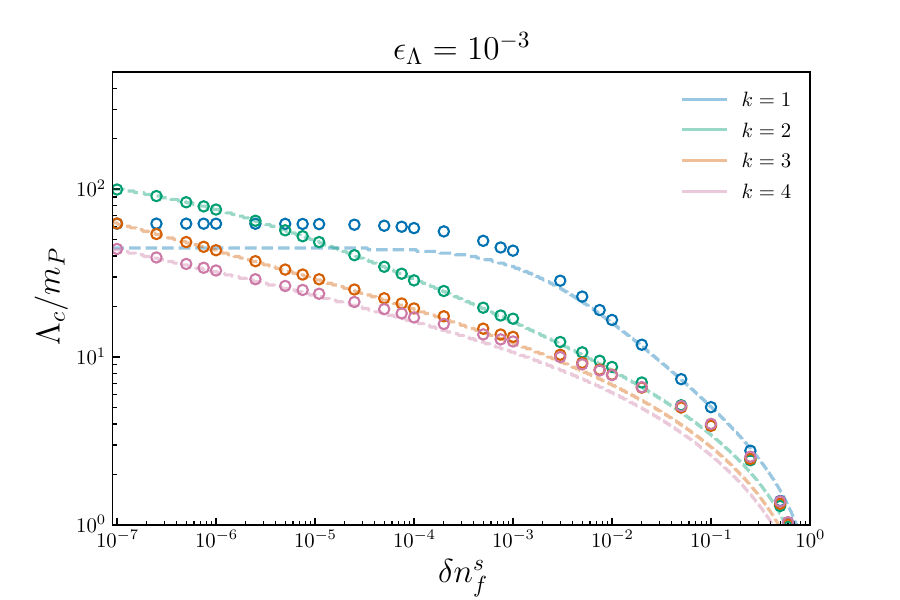}
    \caption{Exact solution (circles) and numerical solution to Eq.~\ref{eq:recursiveFormula} for two values of $\epsilon_\Lambda$. The functions $f(\lambda, m_P^2/Q^2)$ and $g(\lambda, m_P^2/Q^2)$ along with the constant $A$ were chosen so that the expansion of Eq.~\ref{eq:general_running} matches the two-loop $\Psi$ function (Eq.~\ref{eq:two_loop_psi}) to order $\lambda^2$.  See Appendix~\ref{sec:app_numerical_psi} for details. On the left, $\epsilon_\Lambda \gg \epsilon_c$ for most of the plot, so $\Lambda_c$ approaches a constant value as $\epsilon_c \rightarrow 0$. On the right, $\epsilon_\Lambda \ll \epsilon_c$ for most of the plot, and the $(\delta n_f)^{-1/2k}$ scaling is apparent.} \label{fig:confinement_scale_eps_lam}
\end{figure}

If $\epsilon_\Lambda \gg \epsilon_c$, the running of $\lambda$ is primarily due to the region $\lambda(\Lambda^2) < \lambda < \lambda^\star$, and $\Lambda_c$ scales like
\begin{align}
    \Lambda_c \sim \Lambda\left(\frac{m_P}{\sqrt{\epsilon_\Lambda}\Lambda}\right)^{1/(1+k\epsilon_c^{k-1}f(\lambda^\star,0)\lambda^{\star2})}
\end{align}
In particular, as $n_f \rightarrow n_f^s$,
\begin{align}
    \Lambda_c = f(\lambda^\star,0)\lambda^{\star2}g^{(0,1)}(\lambda^\star,0)\frac{m_P}{\sqrt{\epsilon_\Lambda}}\label{eq:lambda_c_k>1},
\end{align}
or (for $k = 1$)
\begin{align}
    \Lambda_c = \frac{f(\lambda^\star,0)\lambda^{\star2}g^{(0,1)}(\lambda^\star,0)}{1+f(\lambda^\star,0)\lambda^{\star2}}\frac{m_P}{\sqrt{\epsilon_\Lambda}}\label{eq:lambda_c_k=1}.
\end{align}
Here, the standard hyper-scaling appears to be enhanced by a constant factor $1/\sqrt{\epsilon_\Lambda}$, but no longer has strong $n_f$ dependence. The behavior of $\Lambda_c$ for small $\delta n_f$ is shown in Fig.~\ref{fig:confinement_scale_eps_lam} for two representative values of $\epsilon_\Lambda$. The $(\delta n_f)^{1/2k}$ scaling is present for a wider range of $\delta n_f$ when $\epsilon_\Lambda$ is small. Otherwise, the value of $\Lambda_c$ plateaus quickly for small $\delta n_f$, as predicted by Eqs.~\ref{eq:lambda_c_k>1} and \ref{eq:lambda_c_k=1}.

\section{Conclusion} \label{sec:conclusion}

In this paper, we have studied the dependence of the induced confinement scale  $\lambda(\Lambda_c) = \lambda_c$ on the number of fermion flavors $N_f$.  For large $N_f$ near the loss of asymptotic freedom, we have confirmed the known exponential suppression of $\Lambda_c$ with respect to the scale of the fermion mass $m$.  

For $N_f$ approaching the conformal sill $N_f^s$, we have argued above using two different lines of argument that the confinement scale $\Lambda_c$ will obey the scaling relation
\beq
\Lambda_c \propto \frac{m_P}{(N_f - N_f^s)^{\zeta}},
\eeq
where the precise value of the exponent $\zeta$ in the denominator is sensitive to the nature of the conformal transition; in the simplest BKT-like transition scenario, $\zeta = 1/4$, while in general $\zeta = 1/(2k)$ for integer $k$, where $k$ is related to the nature of the $\beta$ or $\Psi$ function at the edge of the conformal transition.  This is written in terms of the physical mass $m_P$; using the identification $m_P \sim M$ as discussed in \ref{ssec:gml} above, we can rewrite this relation in terms of a running mass in the form
\beq
\Lambda_c \propto \frac{M}{(N_f - N_f^s)^{1/4}} \propto \frac{m_0^{1/(1+\gamma^\star)}}{(N_f - N_f^s)^{1/4}}.
\eeq

In other words, even near the conformal sill we expect hyperscaling of the mass spectrum as a function of the fermion mass.  However, due to the sharp enhancement of the confinement scale, we have that for $N_f \searrow N_f^s$ the confinement scale will be much larger than the fermion mass, $m_0 \ll \Lambda_c$.  

Although the parametric dependence of $\Lambda_c$ on the fermion mass remains (as it must within the conformal window, since the theory will not confine at $m = 0$), the existence of a regime where $m_P \ll \Lambda_c$ is very interesting.  It is possible that this separation of scales indicates the opening of a regime within the conformal window in which chiral symmetry breaking will be approximately spontaneous, in the sense that observables related to chiral symmetry breaking will be dominated by contributions from $\Lambda_c$ and direct dependence on $m_P$ will be negligible.  

If this is the case, we may expect that the pion states become near-massless pseudo-Nambu-Goldstone bosons compared to the rest of the spectrum, even within the conformal window - so long as the theory is at small but finite mass.  It would be very interesting to study whether dilaton effective field theories \cite{Matsuzaki:2013eva,Golterman:2016cdd,Golterman:2016lsd,Hansen:2016fri,Appelquist:2017vyy,Appelquist:2017wcg,Golterman:2018mfm} could be extended to apply to mass-deformed theories just inside the sill of the conformal window, and whether this extension could give new qualitative insights for lattice studies or for applications to phenomenology. 
In \cite{Del_Debbio_2022} for instance, a new phase with light dilatons and pions is suggested to exist just below the edge of the conformal window.  Our results may predict that this same qualitative behavior persists on the other side of the conformal transition.  It could be interesting to study the implications of our work for their model in particular.

An interesting direction for future work could be to explore whether the effect described here can also be applied to physical theories other than SU$(N_c)$ Yang-Mills theory.  In particular, $\mathcal{N} = 1$ supersymmetric QCD exhibits a non-Abelian Coulomb phase for a certain range of $N_f$ \cite{Intriligator:1995au}, similar to the conformal window studied here.  It would be worthwhile to study whether a similar enhancement of an induced confinement scale occurs in these theories with mass deformation.

In Fig.~\ref{fig:confinement_scale}, we summarize our results by showing the dependence of the confinement scale $\Lambda_c$ on the number of fermion flavors across the conformal window.  If a similar curve could be obtained numerically from lattice simulations, it may be possible to use the scaling of $\Lambda_c$ to estimate both the edge of the conformal window $N_f^s$ and the nature of the transition via the exponent $k$.  Since most lattice calculations for the conformal window are done at relatively small $N_c$ such as $N_c = 3$, the specific enhancement factor for any given theory is likely to be small, as $n_f = N_f / N_c$ will vary in relatively large and discrete steps.  However, it may be possible to determine the scaling with $N_f$ even if the enhancement of $\Lambda_c$ is not very large for any particular value of $N_f$ simulated.  As discussed in Sec.~\ref{sec:hs_sill} and shown in Fig.~\ref{fig:confinement_scale_eps_lam}, the apparent divergence of $\Lambda_c$ as $N_f^s$ is approached will saturate and cut off depending on the value of the ultraviolet cutoff $\Lambda$ at which the running begins; control of lattice cutoff dependence will thus be an important systematic effect to consider in any lattice study of this phenomenon.

\section*{Acknowledgments}

We thank Hooman Davoudiasl for valuable conversations in the early stages of this work, and Anna Hasenfratz, Yigal Shamir, and Maarten Golterman for helpful discussions of an earlier draft.   This work is supported by the U.S. Department of Energy, Office of Science, Office of High-Energy Physics under Grant Contract DE-SC0010005 (E.~N. and R.~M.).

\bibliography{mass_deformation}

\appendix

\section{The Banks-Zaks Limit in $\MSbar$ \label{sec:app_BZ}}

In the Banks-Zaks limit, we can reliably make use of the two-loop $\beta$-function.  The two-loop universal $\beta$-function takes the form
\beq
\beta(\alpha) = \frac{d\alpha}{d(\log \mu^2)} = -\beta_0 \alpha^2 - \beta_1 \alpha^3 + ...
\eeq
with the coefficients equal to
\begin{align}
\beta_0 &= \frac{1}{4\pi} \left( \frac{11}{3} N_c - \frac{4}{3} T(R) N_f \right), \\
\beta_1 &= \frac{1}{(4\pi)^2} \left( \frac{34}{3} N_c^2 - \left[ \frac{20}{3} N_c + 4C_2(R) \right] T(R) N_f  \right).
\end{align}
Here $T(R)$ is the trace normalization, also known as the first Casimir invariant; we adopt the normalization convention that $T(R) = 1/2$ for the fundamental representation of SU$(N)$.  $C_2(R)$ is the standard quadratic Casimir invariant of representation $R$.

The predicted value of the fixed-point coupling is then
\beq
\alpha^\star = -\frac{\beta_0(N_f)}{\beta_1(N_f)}
=-\frac{1}{N_c}\frac{4\pi(11-2n_f)}{34-13n_f}.
\eeq

In this regime, there is generally a more significant distance between $M$ and $\Lambda_c$ than in the focus main body of the paper.  In \cite{Miransky:1998dh}, Miransky suggests that in the Banks-Zaks case, $\Lambda_c=Me^{-\frac{1}{\frac{11}{6\pi}N_c\alpha^*}}$.  If so, we can now recognize the $\MSbar$ result matches our Gell-Mann-Low calculation from equation \ref{eq: BanksZaksGML}:
\beq
    \Lambda_c= M\exp\left(-\frac{225}{88}\frac{1}{\delta n_f^{AF}}\right).
\eeq

\section{Numerical Evaluation of $\Psi$ Function} \label{sec:app_numerical_psi}
\noindent In Section \ref{ssec:alpha_star_exp}, we consider a model Gell-Mann-Low $\Psi$ function of the form
\begin{align}
    \frac{d\lambda}{d\log{Q^2}} = -p(\lambda, m_P^2/Q^2)\lambda^2\left[(\lambda_c - \lambda)^k - A (n_f - n_f^s) + q(\lambda, m_P^2/Q^2) \right]\label{eq:psi_general}
\end{align}
with the additional constraint that $q(\lambda, 0) = 0$ and $q(\lambda, \infty) = A n_f$. In order to numerically evaluate this differential equation, we need to select reasonable values for $\lambda_c$, $A$, and $n_f^s$, and reasonable functions for $p$ and $q$. Rather than choosing parameters arbitrarily, we require that for small $\lambda$, the differential equation should be equivalent to the two-loop $\Psi$ function
\begin{align}
    \left(\frac{d\lambda}{d\log Q^2}\right)^{(2)} = -\lambda^2\left[\frac{11}{3} - \frac{2}{3}n_f(1-f^{(0)}(m_P^2/Q^2)) + \left(\frac{34}{3}-\frac{13}{3}n_f(1 - f^{(1)}(m_P^2/Q^2))\right)\lambda\right]
\end{align}
with $f^{(i)}(0) = 0$ and $f^{(i)}(\infty) = 1$. Let us begin by matching at $m_P^2/Q^2 = 0$: 
\begin{align}
    \frac{d\lambda}{d\log{Q^2}} &=
    -\lambda^2\left[p^{(0)}(0)(\lambda_c^k + An_f^s - An_f) + (p^{(0)}(0)k \lambda_c^{k-1} + p^{(1)}(0)(\lambda_c^k + A n_f^s - An_f))\lambda\right].
\end{align}
For simplicity, we assume that $p$ is independent of $n_f$, so that matching can be done term-by-term in $n_f$. Matching $n_f$ and $\lambda$ dependence, we have the following equations
\begin{align}
    &p^{(0)}(0)(\lambda_c^k + An_f^s) = \frac{11}{3} && p^{(0)}(0)k\lambda_c^{k-1} + p^{(1)}(0)(\lambda_c^k + An_f^s) = \frac{34}{3} \\
    &p^{(0)}(0)A  = \frac{2}{3} && p^{(1)}(0)A = \frac{13}{3}
\end{align}
These equations can be combined to yield
\begin{align}
    \lambda_c^k = A(n_f^{AF} - n_f^s)\label{eq:alpha_c}
\intertext{and}
3 k\lambda_c^{k-1} + 26A n_f^{AF}= 68A
\end{align}
where $n_f^{AF} = 11/2$.
Noting that $34 - 13n_f^{AF}=75/2$ and multiplying by $\lambda_c$ allows us to substitute Eq. \ref{eq:alpha_c}, and we find
\begin{align}
    \lambda_c &= \frac{4 k}{75}(n_f^{AF} - n_f^s).
\end{align}
Using this, we can eliminate $\lambda_c$ from Eq. \ref{eq:alpha_c}, yielding
\begin{align}
    A = \left(\frac{16\pi k}{75}\right)^k(n_f^{AF} - n_f^s)^{k-1}.
\end{align}
Hence, with this model $\Psi$ function, $A$ and $\lambda_c$ can be entirely determined in terms of $k$ and $n_f^s$. For the numerical simulations, we choose various $k$ and $n_f^s = 4$. We note that when $k = 2$ (which is the simplest BKT-like model) and $n_f^s = 4$ (which is supported by theoretical considerations \cite{Appelquist:1998rb}), one finds $4\pi\lambda_c = 16\pi/25 \approx 2.01$. This is very close to the large-$N_c$ critical coupling $N_c\alpha_c = 2\pi/3\approx 2.09$ predicted by analyzing the gap equation.\cite{Appelquist:1998rb}

Let us now match at $m_P^2/Q^2 \rightarrow \infty$. The first two formulae are virtually the same, with $p^{(i)}(0)$ replaced with $p^{(i)}(\infty)$, and there are no other formulae since there is no more $n_f$ dependence. This implies that $p^{(i)}(0) = p^{(i)}(\infty)$. Let us additionally assume that $p$ is completely constant in $m_P^2/Q^2$, so that the only additional running comes from $q$. The $Q$-dependent terms are then of the form
\begin{align}
    \left(\frac{d\lambda}{d\log Q^2}\right)_Q = -\lambda^2[p^{(0)}q^{(0)}(m_P^2/Q^2) + [p^{(1)}q^{(0)}(m_P^2/Q^2)+p^{(0)}q^{(1)}(m_P^2/Q^2)]\lambda ]
\end{align}
Immediately implying that
\begin{align}
    p^{(0)}q^{(0)}(m_P^2/Q^2) &= \frac{2}{3} n_f f^{(0)}(m_P^2/Q^2)\\
    p^{(1)}q^{(0)}(m_P^2/Q^2) + p^{(0)}q^{(1)}(m_P^2/Q^2) &= \frac{13}{3}n_ff^{(1)}(m_P^2/Q^2)
\end{align}
We can multiply these equations on both sides by $A$ to get rid of the $p^{(i)}$ to find that
\begin{align}
    q^{(0)}(m_P^2/Q^2) &= An_f f^{(0)}(m_P^2/Q^2)\\
    q^{(1)}(m_P^2/Q^2) &= \frac{13}{3}An_f(f^{(1)}(m_P^2/Q^2) - f^{(0)}(m_P^2/Q^2)).
\end{align}
These are the functions and parameters we choose for the numerical solutions in Fig.~\ref{fig:confinement_scale_eps_lam}. Although Eq.~\ref{eq:psi_general} is slightly more general than Eq.~\ref{eq:general_running} in the text, the results we derive in the text are consistent for small $\delta n_f$ and small $m_P^2/Q^2$ as predicted.


\end{document}